\documentclass[aps,twocolumn,prd,epsf,showpacs,preprintnumbers]{revtex4}

\usepackage{amssymb}
\usepackage{graphicx}

\newcommand{\be}{\begin{equation}}
\newcommand{\ee}{\end{equation}}
\newcommand{\bea}{\begin{eqnarray}}
\newcommand{\eea}{\end{eqnarray}}

\newcommand{\ch}{{\cal H}}

\begin{document}
%\begin{titlepage}

%\vspace{1in}

\preprint{AEI--2004--027}

\title{Inflationary Cosmology and Quantization Ambiguities\\
in Semi-Classical Loop Quantum Gravity}

\author{Martin Bojowald$^1$, James E. Lidsey$^2$, David J. Mulryne$^2$,
Parampreet Singh$^3$ and Reza Tavakol$^2$}

\affiliation{$^1$Max-Planck-Institut f\"ur Gravitationsphysik, 
Albert-Einstein-Institut, 14476 Potsdam, Germany}

\affiliation{$^2$Astronomy Unit, School of Mathematical Sciences,  
Queen Mary, University of London, Mile End Road, LONDON, E1~4NS, UK}

\affiliation{$^3$IUCAA, Ganeshkhind, Pune 411 007, India} 

\begin{abstract}
In loop quantum gravity, modifications to the geometrical density 
cause a self-interacting scalar field to accelerate
away from a minimum of its potential. In principle, this  
mechanism can generate the conditions that subsequently lead to  
slow-roll inflation. The consequences for this mechanism 
of various quantization ambiguities arising within loop quantum cosmology
are considered. For the case of a quadratic potential, 
it is found that some quantization
procedures are more likely to generate a phase of 
slow--roll inflation. In general, however,  
loop quantum cosmology is robust to 
ambiguities in the quantization and extends the 
range of initial conditions for inflation. 
\end{abstract}

\pacs{98.80.Cq,04.60.Pp}
 
\maketitle
 
%\end{titlepage}

%\double

\section{Introduction}
\setcounter{equation}{0}

Loop quantum gravity (LQG) or quantum geometry  is at present the main
background independent and non--perturbative
candidate for a quantum theory of gravity (see for example 
\cite{rovelli98,thiemann02}).
Key successes of this approach have been the prediction of discrete
spectra for geometrical operators \cite{geometrical_op},
the existence of well defined
operators for the matter Hamiltonians which
provides a cure for the ultraviolet divergences
\cite{thiemann_matter}, and the derivation of 
the Bekenstein--Hawking entropy formula \cite{bek_hawking}.

Given that LQG effects are likely to have important
consequences in high energy and high curvature regimes, 
early universe cosmology 
provides a natural environment to test these new features. 
In recent years, considerable interest has focused on 
applying LQG to minisuperspace cosmological models 
(see \cite{martin_hugo,martin_kevin_2,ICGC} for reviews)
and this has resolved various difficulties
encountered in  conventional 
Wheeler--de Witt quantization for isotropic models \cite{martin1,martin2} and
anisotropic models \cite{martin_date}.

{}From the loop quantum cosmology (LQC) perspective,  
the evolution of the universe is comprised of the three
distinct phases. Initially, there is a truly discrete quantum
phase which is described 
by a difference equation \cite{martin1,martin2}. 
A key consequence of this discretization
is the removal of the initial singularity \cite{martin1}.
As its volume increases, the universe 
enters an intermediate semi--classical phase in which the
evolution equations take a continuous
form but include modifications due to non--perturbative 
quantization effects \cite{martin3}.
Finally, there is the classical phase
in which the usual continuous ODE/PDE cosmological
equations are recovered and quantum effects vanish.

The above intermediate phase (which is demarcated by two scales, as
discussed in detail in \S 2) is the most important as far as
phenomenological studies of LQC are concerned and
will be subject of our study here. Key results arising from this
phase include the setting up of 
suitable initial conditions for inflation \cite{martin3,martin_kevin_1,tsm03},
possible signatures
on the CMB spectrum such as the loss of power at largest
scales and the running of the scalar spectral index \cite{tsm03} and
the avoidance of a big crunch in closed models \cite{st03}.

Despite these successes, however, it is known 
that the quantization procedure itself is not
unique. In general, quantization ambiguities arise when 
composite operators are constructed from the corresponding classical
expressions. Some of these ambiguities are more fundamental 
from a theoretical point of view than others. Nevertheless, 
it is possible that the range of different choices that lead to a viable 
cosmological model might be significantly narrowed
by employing not only theoretical considerations but 
also by confronting phenomenological predictions with observations.
Two immediate questions arise in this connection:
(i) do some/all ambiguities of this kind
lead to observable predictions/consequences and (ii)
are some predictions of LQC (such as establishing
the correct initial conditions for inflation)
robust with respect to these ambiguities?
For example, 
if a particular ambiguity were to result in observational 
effects that were subsequently detected, this 
could provide an observational basis for invoking that particular
choice. On the other hand, the inability of ambiguities to produce
observationally distinguishable consequences would make it difficult to
discriminate between them without theoretical input.
Whatever the outcome, a study of these questions 
would demonstrate the range of
phenomenological possibilities allowed by LQC and would thus 
enable a clearer comparison to be made between the theoretical predictions 
and observations. 

A study of the effects of quantization ambiguities on the LQG's
capacity to remove singularities was recently made
\cite{bojowaldCQG2002}, where it was found that the removal of the
big-bang singularity is robust with respect to a number of different
ambiguities.  Our aim here is to study the effects of quantization
ambiguities on the evolution of the very early universe. In
particular, we shall study the way both kinematical and dynamical
ambiguities can affect the ability of LQC to set up the appropriate
conditions for inflation.  We focus primarily on the possibility that
in LQC inflation may arise naturally even when the inflaton is
initially located near to a minimum of its potential. As we shall see,
there are two mechanisms that affect the setting of the initial
conditions for conventional slow--roll inflation. Firstly, there is an
anti-friction effect induced by the semi-classical modification of the
scalar field equation of motion which can, in an expanding universe,
accelerate the field away from its initial value. Secondly, there are
effects due to the quantum mechanical uncertainty relations. Since
these mechanisms are in principle unrelated, each is treated
separately in order to maintain a clear distinction between the
two. Finally, in phenomenological studies of this nature, there are
constraints, in addition to the observational bounds, that arise from
the requirement that the region of parameter space under consideration
should be consistent with the approximations that apply in the
semi--classical and classical epochs.  We shall see that these also
provide strong constraints on the set of ambiguities that lead to
appropriate initial conditions for the onset of inflation.

The plan of the paper is as follows. In \S 2, the ambiguities that 
arise in LQC are discussed, including a comparison 
of such ambiguities from the perspective of the 
full theory. Section 3 considers issues relevant to the setting of 
initial conditions in inflation. Section 4 presents an approximate 
analytical scheme for following the evolution of the field 
during the semi--classical phase and \S 5 presents the numerical results. 
We conclude with a discussion and consider future directions in \S 6.

%---------------------------------------------
\section{Quantization ambiguities in loop quantum cosmology}
%----------------------------------------------

There are various freedoms in 
any quantization scheme. In particular for LQG a number of ambiguities arise in the
derivation of the quantum evolution and of the effective cosmological
field equations from the quantum difference equations. In fact,
quantum cosmological field equations require the appearance of
operators that are not fundamental in a loop quantization, but are
composite expressions constructed from basic ones in more or less
complicated ways. In general, a quantization of composite objects does
not have a unique realization and possible consequences of
different choices have to be studied.  The operators affect the
semi-classical behavior which can be modeled by considering their
expectation values in the semi-classical domain, and as we will see it
matters which choice of quantum operator is taken. Among the wealth of
ambiguities there are some with the strongest influence on the
effective classical behavior, which we will discuss in this Section in
more detail and then use in the rest of the paper.

We focus on the spatially isotropic and topologically compact
Friedmann--Robertson--Walker (FRW) models sourced by a single scalar
field, $\phi$, that is minimally coupled to Einstein gravity and
self--interacting through a potential $V(\phi)$. The classical
Hamiltonian for this matter degree of freedom is
\be \label{ham}
\ch_{\phi} =  \frac{1}{2} \, \frac{1}{a^3} \, p_{\phi}^2 + a^3 \, V(\phi)
~,
\ee
where $a$ is the scale factor of the universe and classically
$p_{\phi} = a^3 \dot{\phi}$ is the momentum canonically conjugate to
$\phi$.

In the Wheeler--DeWitt quantization procedure, the Hamiltonian operator
diverges in the limit $a \rightarrow 0$. LQC provides major
insight into this issue. Near $a = 0$ the concept of spacetime does not
exist and one is in a full quantum gravity domain.
The information about spacetime is encoded in the quantum states (spin
networks) on which geometrical operators have discrete
spectra. Moreover, one can quantize even inverse powers of metrical
expressions, which diverge classically, and obtain
well-defined results \cite{thiemann_matter}. In the cosmological context this
implies that eigenvalues of inverse powers of the scale factor
also have a finite bounded spectrum \cite{martin1}.
At larger volumes, where the universe is in a semi--classical state, 
the quantum behavior can be approximated by a
continuous spacetime which retains some of the quantum geometric properties. 
It is these semi--classical effects that provide potentially 
observable consequences.

\subsection{Ambiguity parameters}

\begin{figure}
\includegraphics[width=8.5cm]{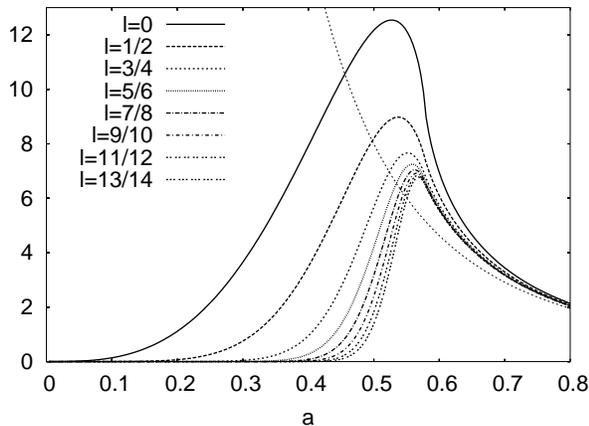}%
\caption{\label{figure1.eps}
Small-$a$ behavior of the effective geometrical density ($d_{j,l}(a)$) for different
values of $l$ and $j=1$. The monotonically increasing dotted line 
corresponds to the classical expectation $a^{-3}$.}
\end{figure}

Since the main reason for the occurrence of classical singularities
is the diverging matter Hamiltonian, it is not surprising that the main
ambiguities that are relevant for our purposes are contained in the
quantization of the geometrical density, $d(a):=a^{-3}$, that is present in the
kinetic term of the matter Hamiltonian.  After quantization,
the discrete spectrum of the geometrical density can be approximated as a 
continuous function of $a$ carrying quantum
effects. Its precise form can be computed once a particular
quantization scheme is chosen, resulting in an expression
characterized by two parameters, $\{ j, l \}$ \cite{bojowaldCQG2002,ICGC}:
\be \label{dj}
d_{j,l}(a)=D_l(q)a^{-3}\,,~q=a^2/a_*^2\,,~ a_*^2=\gamma l_{\rm Pl}^2\,j/3 \, ,
\ee
where $\gamma \approx 0.13$ is the
Barbero-Immirzi parameter and $l_{\rm Pl}$ is the Planck length
(see Fig.\ \ref{figure1.eps}) .  As
we will see, the scale $a_*$ determines the size of the scale factor
below which the geometrical density is significantly different from
its classical form, which cannot be captured even perturbatively on
smaller scales, $a \lesssim a_*$. The size of $a_*$ is determined by
the half integer $j$, which is our first ambiguity parameter, and is
so far unrestricted by considerations in the full theory (even though
one can argue that smaller values appear more natural; see the
following subsection). The function $D_l$ characterizing the modified
density is subject to further ambiguities. Here we only focus on one
parameter $l$, which determines the behavior of the density on scales
small compared to $a_*$ \cite{ICGC}. The parameter $l$ can take any
value between zero and one, and leads to the function

\begin{eqnarray} \label{defD}
 D_l(q) =
\left\{\frac{3}{2l}q^{1-l}\left[(l+2)^{-1}
\left((q+1)^{l+2}-|q-1|^{l+2}\right)\right.\right. \nonumber\\
 - \left.\left.\frac{1}{1 + l}q
\left((q+1)^{l+1}-{\rm
sgn}(q-1) 
|q-1|^{l+1}\right)\right]\right\}^{3/(2-2l)} . 
\end{eqnarray}

Common features for all values of $j$ and $l$ are that $d_{j,l}(a)$
approaches the classical behavior $a^{-3}$ for $a\gg a_*$, that there
is a peak value around $a_*$ where the density is maximal, and that
for $a\ll a_*$ the density approaches zero in a power-law behavior
whose exponent, $3/(1-l)$, is determined by $l$:
\be
\label{approxD} d_{j,l}(a) \sim
(3/(1+l))^{3/(2-2l)}(a/a_*)^{3(2-l)/(1-l)} a^{-3} ~.
 \ee

So far we have only discussed the kinematical properties of the density
operator, which already give us a taste of the quantum properties to
expect.  Since loop quantum gravity is a canonical quantization, its
dynamics are encoded in the Hamiltonian constraint equation which in
the cosmological context is a difference equation for the wave
function \cite{bojowald02a}. It is a quantization of the classical expression
\begin{equation}\label{constraint}
 {\cal H}:=-3\dot{a}^2a+8\pi G {\cal H}_{\phi}=0
\end{equation}
which, upon dividing by $a^3$, is nothing but the Friedmann
equation. From the point of view of the quantization, however, the
primary object is ${\cal H}$; there is no direct
quantization of the Friedmann equation. Moreover, the classical
constraint ${\cal H}$ plays the role of the Hamiltonian of the whole
system, gravity plus matter, which determines the full dynamics via
the Hamiltonian equations of motion. For the matter part, this results
in equations
\[
 \dot{\phi}=\{\phi,{\cal H}\} \quad,\quad
\dot{p}_{\phi}=\{p_{\phi},{\cal H}\}
\]
which can be combined to form a second order differential equation for
$\phi$, the scalar field equation. In addition to the Friedmann
equation, there is also a second order equation for the scale factor,
the Raychaudhury equation, which is obtained via a Poisson bracket of
the gravitational degrees of freedom (but follows also from the
continuity equation for the matter together with the Friedmann
equation).

We can now look at how quantum effects can occur in an
effective, semi-classical expression of the constraint. For scales above
\begin{equation}
\label{initialscaledef}
a_i \approx \sqrt{\gamma} l_{\rm Pl},
\end{equation}
\noindent we assume the spacetime can be approximated
by a continuous manifold. Since the full quantum operator is a
difference operator, the discreteness will lead to perturbative
corrections of higher order in $\dot{a}$, but they only play a role
close to the scale $a_i$.  More important is the presence of the
modified geometrical density in the kinetic term of the matter
Hamiltonian.  The main modification of the constraint is obtained,
therefore, by replacing $a^{-3}$ in the matter Hamiltonian with the
modified density $d_{j,l}(a)$. In this way, the effective
semi-classical dynamics from the constraint
\begin{equation}\label{effconstraint}
 {\cal H}:=-3\dot{a}^2a+8\pi G \langle\hat{{\cal H}}_{\phi}\rangle=0
\end{equation}
can also become sensitive to the same quantization ambiguities as the matter
Hamiltonian. For a scalar field, we have
\begin{equation}
 {\cal H}:=-3\dot{a}^2a+8\pi G (d_{j,l}(a)p_{\phi}^2/2+a^3 V(\phi))=0\,.
\end{equation}

So we see that for our study there are two different scales which are
important.  The first is defined by $a_i$, above which we assume a
classical spacetime, but below which the full difference equations must
be employed.  The second scale is defined by $a_*$, below which
the modifications to the behavior of the geometric density become important.
We see that when $j$ is larger than three, there is an overlap between the
two scales, so spacetime can be considered as continuous, but quantum effects
in the density are still important.  This region of overlap then
defines the semi-classical regime in which our study is based.

After dividing by $a^{-3}$ (which will not be modified since it is
just a classical manipulation), we obtain the effective Friedmann
equation
\be \label{fredeq1}
H^2 = \frac{8\pi {l_{\rm Pl}^2}}{3} \, \frac{1}{a^3} \, \langle \hat
\ch_{\phi}
\rangle
\ee
which for a scalar field is of the form
\be \label{fredeq1a}
H^2 = \frac{8\pi l_{\rm Pl}^2}{3} \left[ \frac{1}{2} \, D_l^{-1}\, \dot{\phi}^2
+
V(\phi) \right] \, ~.
\ee
Here, the scalar field momentum,
\be
\label{pdiff}
 p_{\phi}=a^3D_l^{-1}\dot{\phi}
\ee
is different from the classical momentum due to the modified matter
Hamiltonian and Eq. (\ref{pdiff}) 
follows from the Hamiltonian equation of motion for
$\phi$, $\dot{\phi}= \partial {\cal{H}} /\partial
p_{\phi}$. The effective scalar field equation is then derived from the
second Hamiltonian equation, $\dot{p}_{\phi}=- \partial {\cal{H}}/ \partial
\phi$,
and takes the form 
\be
\ddot{\phi} +\left( 3H - \frac{\dot{D_l}}{D_l} \right) \dot{\phi} +
D_l V'(\phi) =0\,.
\label{scalareom}
\ee

This derivation of the effective classical equation is the original one of
\cite{martin3, martin_kevin_1}, and we will refer to it as {\sc Ham}
since here the Hamiltonian is the primary dynamical object.

At this point there is the possibility of additional ambiguities since
we can now consider the full Hamiltonian as a composite object. It is, for
instance, possible to insert arbitrary positive powers of
$a^3d_{j,l}(a)$ into the expression because this 
factor would just be equal to unity at the classical level. 
In particular, we can modify the right hand side of the
Friedmann equation by multiplying the matter Hamiltonian with such a
power,
\be \label{Friedn}
H^2 = \frac{8\pi l_{\rm Pl}^2}{3} D_l^n\left[ \frac{1}{2} \,
D_l\,a^{-6} p_{\phi}^2 + V(\phi) \right]
\ee
where $n>0$ is a new ambiguity parameter. In the same way as before we
can then compute the scalar equations of motion, where now
$\dot{\phi}=a^{-3}D_l^{1+n}p_{\phi}$. Thus,
\be
H^2 = \frac{8\pi l_{\rm Pl}^2}{3} \left[ \frac{1}{2} \, D_l^{-1-n}\,
\dot{\phi}^2 + D_l^n V(\phi) \right]
\ee
and
\be
 \ddot{\phi} +\left( 3H - (1+n)\frac{\dot{D_l}}{D_l} \right) \dot{\phi} +
D_l^{1+2n} V'(\phi) =0\,.
\ee
This more general scheme will be called {\sc Ham}(n) in what follows. For
$n=0$, the equations reduce to those of {\sc Ham}.

For $n=1$, on the other hand, the Friedmann equation (\ref{Friedn})
could also have been obtained by replacing both the $a^{-3}$ in
the kinetic term and the $a^{-3}$ explicit in $H^2= 8\pi l_{\rm
Pl}^2 a^{-3}{\cal H}_{\phi}/3$ with $d_{j,l}$. When starting with the
Hamiltonian constraint, however, this replacement would happen only at
the matter side but not in the gravitational part $H^2\equiv
a^{-3}\dot{a}^2a$ obtained from (\ref{constraint}) by dividing by
$a^3$.

An alternative to this approach has been advocated in 
which the Hubble parameter is viewed as the primary
object in order to derive the effective dynamical field equations
\cite{golam}.
While the Hubble operator still involves the matter Hamiltonian, there
is an additional difference in its spectrum arising due to
a quantization of the $1/a^3$ term in (\ref{fredeq1}) which
thus can be replaced by
\be \label{fredeq2}
H^2 = \langle \hat H^2 \rangle = \frac{8\pi}{3} \, {l_{\rm Pl}^2} \,
\langle \hat{a^{-3}} \rangle \,
 \langle \hat \ch_{\phi} \rangle
\ee
This results in a Friedmann equation different in form to that of
Eq. (\ref{fredeq1a}), but identical to that of {\sc Ham}(1) in terms
of $p_{\phi}$:
\be
H^2 = \frac{8\pi l_{\rm Pl}^2}{3} \left[ \frac{1}{2} D_l^2 a^{-6}
p_{\phi}^2 + D_l \, V(\phi) \right] \, ~.
\ee

In contrast to the scheme {\sc Ham}(1), however, it is implicit in
\cite{golam} that the scalar momentum is not changed compared to
{\sc Ham}. As a consequence, the scalar field equation still has the
form of {\sc Ham}(0) while the Friedmann equation in terms of
$p_{\phi}$ is that of {\sc Ham}(1). In terms of $\dot{\phi}$, however,
the Friedmann equation has a new form,
\be \label{fredeq2a}
H^2 = \frac{8\pi l_{\rm Pl}^2}{3}  \left[ \frac{1}{2} \dot{\phi}^2 +
D_l \, V(\phi) \right]
\ee
since the relation between $\dot{\phi}$ and $p_{\phi}$ is different.
We will refer to this alternative way of obtaining the Hubble equation
as {\sc Fried} since the dynamical law is obtained from a form
analogous to the Friedmann equation.

The quantum theory provides further freedom for deriving Friedmann
equations and results in further ambiguity. We just mention one other
example, namely that of writing the Hamiltonian in Eq. (\ref{ham}) in
a classically equivalent form as
\bea \label{ham1}
\ch_{\phi} &=&  \frac{1}{2} \, \frac{1}{a^3} \, p_{\phi}^2 + a^3 \,
V(\phi) ~ \nonumber \\
&=& \frac{1}{2} \, \frac{1}{a^{3(n + 1)}} \, a^{3n} \, p_{\phi}^2
+ \frac{1}{a^{3m}} \, a^{3(m + 1)}\, V(\phi)
\eea
where $\{ n, m \}$ are arbitrary, semi--positive definite constants,
$\{ m,n \} \geq 0$ (for $m=n$ we obtain {\sc Ham}(n)). The {\sc Ham} Hubble
equation (\ref{fredeq1}) thus yields
\be \label{fredeq1nm}
H^2 = \frac{8\pi}{3} \, {l_{\rm Pl}^2} \, \left[ \frac{1}{2} \, D_l^{-(n + 1)} \,
\dot{\phi}^2 + D_l^m \, V(\phi) \right] \, ~,
\ee
whereas adopting the {\sc Fried} quantization procedure beginning with
Eq. (\ref{fredeq2}) implies that
\be \label{fredeq2nm}
H^2 = \frac{8\pi}{3} \, {l_{\rm Pl}^2} \, \left[ \frac{1}{2} \, D_l^{-n} \,
\dot{\phi}^2 + D_l^{(m + 1)} \, V(\phi) \right] \, .
\ee
In these extended schemes
the scalar field equation becomes
 \be \label{field}
\ddot{\phi} + \left( 3H - (n + 1) \frac{\dot{D}_l}{D_l} \right) \dot{\phi}
 + D_l^{(m + n + 1)} \,
 V'(\phi) =0 ~.
\label{kgeq_nm}
 \ee

Thus, freedom at the quantum level, as parametrized by non--zero values of
$\{ m, n\}$, results in effective cosmological field equations
that are radically different from the natural (minimal)
choice corresponding to $m=n=0$.

In summary, we have briefly discussed the following quantization
ambiguities: firstly, there is the half-integer $j$ which determines
the position of the peak of the effective density and, consequently,
the value of the scale factor at which classical physics is recovered.
Secondly, we have a parameter $0<l<1$ which specifies the initial
increase of the effective geometrical density. How these parameters
appear from the point of view of full loop quantum gravity will be
discussed in the following subsection.  Moreover, the procedure for
deriving the quantized version of the Friedmann equation is not
unique: differences arise depending upon whether one quantizes the
matter Hamiltonian \cite{martin3} or directly quantizes the Friedmann
equation by viewing the Hubble parameter as an operator
\cite{golam}. Finally, there is an additional freedom in writing down the
Hamiltonian, and hence the Friedmann equation. This is parametrized by
the constants $m$ and $n$.

\subsection{Comparing ambiguities from the point of view of the full
theory}

The ambiguities listed in the preceding discussion do not all appear
at the same level from a theoretical perspective. For instance, the
parameters $j$ and $l$ emerge when we quantize the geometrical density
and are therefore already present at the kinematic level. More
specifically, we have to choose a way to write the classically
divergent $a^{-3}$ such that it is suitable for a quantization. It is
not possible to use the scale factor operator directly since it does
not have an inverse in the loop representation
\cite{InvScale,Bohr}. However, using the canonical pair
$(a^2,p_{a^2})$ and the loop variable
$h_I=\exp(3p_{a^2}\tau_I/8\pi\gamma G)$ with Pauli matrices $\tau_i$,
we can write
\begin{eqnarray} \label{dens}
 a^{-3}&=&\Biggl(3(8\pi\gamma
 Glj(j+1)(2j+1))^{-1}\nonumber\\
 &&\left.\times \sum_I{\rm tr}_j(\Lambda_I^i\tau_i
 h_I\{h_I^{-1},a^{2l}\})\right)^{3/(2-2l)}
\end{eqnarray}
which is a classical identity and involves only positive powers of the
scale factor for $0<l<1$.  The basic quantities for the loop
quantization are the squared scale factor $a^2$ and its conjugate
momentum $p_{a^2}$ appearing in $h_I$. The trace is taken after
evaluating the expression in the irreducible SU(2)-representation with
spin $j$, which is our first ambiguity parameter. In addition there is
the parameter $l$ specifying the power of $a^2$ in the expression for
$a^{-3}$. Both parameters are common to all the other schemes
discussed above. Additional freedom then arises when we quantize the
Friedmann equation, corresponding to ambiguities of the dynamics.

Another difference between the types of ambiguities is that not all of
them appear equally natural, and some values of the parameters can be
preferred compared to others. Thus, a choice may be made already from
a purely theoretical point of view by comparing the expressions in
loop quantum cosmology and the corresponding ones in full loop
quantum gravity.

For the parameter $j$ there are virtually no internal restrictions,
not even when we use the full theory since there the same freedom
appears. Because it corresponds to choosing a non-trivial irreducible
representation, one may argue that the most natural value 
is $j=1/2$ for the fundamental
representation. This is also the smallest allowed value for 
$j$. (An additional argument is that from the fundamental perspective we really
choose two representations, one for the gravitational part of the
constraint and one for the matter part. The gravitational part gives
us the $\dot{a}^2$ which no longer depends on the representation.
Still, one would regard a choice as more natural if the two
representations are close to each other, which points toward smaller
$j$.) But once one does not restrict oneself to this choice, there is
no distinction between higher values except that huge ones (of the
order $10^{20}$ or larger) would be excluded by particle physics
experiments. In this paper we thus keep $j$ as the main ambiguity
parameter to be restricted with observational input. An interesting
question is whether the cosmological evolution of the early universe
would also point to smaller values of $j$ or require it to be very
large.

Concerning the parameter $l$, the situation is different. It occurs
because $a^{-3}$ is rewritten as a Poisson bracket involving a
positive power of $a^2$. From the point of view of the full theory,
$a^2$ corresponds to the isotropic component of a densitized triad, on
which the loop quantization is based. Without symmetry assumptions, it
is much more difficult to rewrite the geometrical density in a way
which is both suitable for a quantization and still covariant under
coordinate transformations. That this is possible had been
demonstrated in \cite{thiemann_matter}, even before the cosmological
calculations had been performed. In this case, not all values $0<l<1$ can
be used and some are preferred, even though there is still no
unique choice. But instead of a continuous range only a discrete
sequence, $l_k=1-(2k)^{-1}$, $k\in{\mathbb N}$ appears (see
Appendix). Moreover, $l_2=3/4$ appears most natural and also
corresponds to the quantization of matter Hamiltonians in
\cite{thiemann_matter}. It would result in an initial power law of the
form $a^{12}$, or in general $a^{6k}$ with $l_k$, for the density.

The schemes {\sc Ham}(n) and {\sc Fried} can also be distinguished by
internal considerations. As explained before, the Hamiltonian is the
primary object in a canonical quantization so {\sc Fried},
which puts the emphasis on the Hubble parameter, is more specific. In
fact, a corresponding quantization in the full theory is not possible,
while the quantization steps of {\sc Ham} are modeled on those of full
loop quantum gravity. Among the different possibilities in {\sc
Ham}(n), it is clear that {\sc Ham}={\sc Ham}(0) is most natural since
it corresponds to no additional insertion of $a^3d(a)$.

To summarize, for $j$ there are the weakest restrictions from the
theoretical side alone, even though small values look more
natural. For the other kinematical parameter $l$ there is a discrete
set of preferred choices when we compare with the full theory, such
that all values lie in the interval $1/2\leq l<1$.  Theoretical
considerations for the dynamical ambiguities strongly prefer the
original scheme {\sc Ham}. The reason for the fact that the dynamical
ambiguities are much more restricted conceptually can be seen in the
different way they emerge. We can not avoid the kinematical
ambiguities since the most direct way to quantize $a^{-3}$ is ruled
out by the non-existence of an inverse of $\hat{a}$ in the loop
quantization. We then have to use a more complicated quantization
obtained after rewriting the classical expression. This opens the door
for ambiguities which are unavoidable. For the Hamiltonian constraint,
on the other hand, the most direct quantization does work and leads us
to {\sc Ham}. The other choices change this procedure in a similar way
to that of the kinematical ambiguities, but these changes are no
longer forced upon us. Thus, the most direct procedure which works
appears as the most natural one.

In the remainder of this paper we present the first investigation of the
ambiguities from the point of view of cosmological phenomenology. In
particular, we are interested in whether or not the conceptual
expectations discussed so far are also favoured by the phenomenological
ones. The main focus will be on the parameter $j$, in particular
contrasting small with large values, and on differences between the
schemes {\sc Ham} and {\sc Fried}.

%-----------------------------
\section{Initial conditions}
%-------------------------------

In this Section we study the phenomenological consequences of various
quantization ambiguities discussed in the preceding Section. Our
main focus will be to determine the importance of these ambiguities
in establishing appropriate initial conditions for
slow--roll inflation.  Within the context of chaotic
inflation, it has been argued that the universe emerges from a spacetime
foam at the Planck scale, where the energy density is of the order
$m_{\rm Pl}^4$ \cite{linde}. The value of the inflaton takes
different values in the different regions of the universe and
inflation proceeds in those regions where the
field has suitable initial values. In the case of a quadratic potential,
$V(\phi ) = m^2 \phi^2 /2$, where $m$ is the mass of the inflaton,
inflation is possible in those regions where $l_{\rm Pl} \phi_i
> 1/4$. (For a review, see, e.g., Ref. \cite{lidlyth}.)

Inflation is presently the most
favored scenario for describing the very early history of the universe
and has received strong support from recent observations
of the CMB power spectrum \cite{wmap}. However,
the question of whether a given set of initial conditions is favored
is difficult to quantify in the absence of a full theory
of quantum gravity. In view of these developments,
it is important to investigate physical processes that
enable the inflaton field to reach the values required for
inflationary expansion.

Recent developments in loop quantum cosmology have
provided a mechanism for setting up the necessary conditions
for inflation even if the field is initially located in a
minimum of its potential and has a low kinetic energy
\cite{martin3,martin_kevin_1,tsm03}.
At sufficiently small volume $(a_i < a \ll a_*)$, 
quantum mechanical
effects cause the universe to undergo a superinflationary
expansion, $\dot{H}>0$, which is not driven by the potential energy of the
inflaton \cite{martin3}. The asymptotic form 
of $D_l$ implies that the frictional term in the
scalar field equation (\ref{scalareom})
changes sign. Hence, the expansion of the
universe acts as a driving term and accelerates the
field away from the potential minimum.
Since the potential term in the scalar field equation
(\ref{scalareom}) is negligible for $D_l \ll 1$, the
kinetic energy of the field rapidly dominates the
dynamics. 

Once the universe has expanded sufficiently ($a > a_*$), the
conventional classical dynamics is recovered.
The field decelerates, reaches
a maximum displacement and rolls back down the potential. If
the field is able to move sufficiently far up its potential, the
conditions relevant to standard, slow--roll inflation may be realized in
a natural way. Moreover, if the field
reaches its point of maximal displacement some 60
e-foldings or so before the end of inflation, the perturbations
generated during the turning point could lead to 
observable effects in the CMB \cite{tsm03}.  

{}From a conceptual point of view, this suggests that the
set of initial conditions that leads to slow--roll
inflation might be significantly widened in loop quantum cosmology.
However, a crucial question that must be addressed is
whether this behaviour is robust under the
quantum ambiguities discussed in \S 2.
The scalar field equation (\ref{scalareom})
has the same functional form for both the {\sc Ham} and {\sc Fried}
quantization schemes and, since $D_l
\ll 1$ for $a < a_*$, we expect 
the universe to rapidly enter an epoch of superinflation
in both cases.

For a more quantitative analysis we begin at the onset of the
semi--classical regime $(a \approx a_i)$, where we can approximate the
difference equations as coupled ODEs represented by the modified
Friedmann and scalar field equations for each quantization scheme.
An immediate question that arises is
the range of appropriate values for the
initial conditions $\{ \phi_i , \dot{\phi}_i \}$. If one views the
inflaton as a localized wavepacket, this has a spread in position and
velocity. Such a spread has a lower bound due to the minimal
uncertainty relation, $|\Delta \phi \Delta p_{\phi}| \ge 1$.  
For the extended
quantization scheme of Eq. (\ref{ham1}), the momentum
canonically conjugate to the field is given by $p_{\phi} =
(a^3d_{j,l})^{-(n+1)} a^3 \dot{\phi}$, and so the uncertainty 
relation can be written as:
\begin{equation}
\label{uncertain}
\left| \Delta \phi \Delta \dot{\phi} \right| 
\ge \left[ d_{j,l} (a_i) \right ]^{n+1} a_i^{3n} ~.
\end{equation}
This is equivalent
to considering a semi-classical state of the inflaton when it emerges
into the classical regime. To proceed further, one has to understand how
the transition from a wave packet to sharp, classical initial
conditions is made. In general, this issue involves the
interpretation of the wave function in quantum cosmology and the
measurement process and is beyond the scope of the present paper. In view
of this, we keep our assumptions as weak as possible
but take into account the essential effects arising from the
uncertainty principle.

Since the uncertainty principle only limits quantum physical
fluctuations, and not the expectation values, it would certainly be
consistent to specify $\phi_i=0=\dot{\phi}_i$ initially.
However, such an assumption would effectively ignore the
uncertainty principle and any possible effects originating from
fluctuations of the mean inflaton value. Indeed, a standard probabilistic
interpretation would imply that the most likely
values for the inflaton are of the order of the
spreads $\Delta\phi$ and $\Delta\dot\phi$, whereas very
small values would be unlikely.
As order of magnitude estimates are employed 
in what follows, we will identify the initial
velocity of the inflaton with the spread in velocity $(\dot \phi_i \sim
\Delta \dot \phi)$.  Eq. (\ref{uncertain}) then provides a measure of
the uncertainty in the field's position on the potential ($\phi_i
\approx \Delta \phi$, assuming $\phi_i =0$ classically). This
provides a measure for choosing $\{\phi_i , \dot{\phi}_i \}$ as
initial conditions in the ODE's.

Using the $a \ll a_*$ limit of $d_{j,l}$ given in Eq. (\ref{approxD}), 
Eq. (\ref{uncertain}) implies that
\begin{equation}
\left| \phi_i \dot{\phi}_i \right| > \left[ \left( \frac{3}{1+l}
\right)^{3/(2-2l)} \left( \frac{a_i}{a_*} \right)^{3(2-l)/(1-l)}
\right]^{n+1} a_i^{-3}
\end{equation}
The discrete nature of spacetime becomes significant on scales below
$a_i \approx \sqrt{\gamma} l_{\rm Pl}$ and it is natural to
invoke this estimate as the initial value of the scale factor. In this
case, and with $l=3/4$,
\begin{equation}
\label{uncert}
\left| \phi_i \dot{\phi}_i \right| \ge 10^{5n+6.31} j^{-15(n+1)/2}
l_{\rm Pl}^{-3}   .
\end{equation}
For $n=0$, Eq. (\ref{uncert}) simplifies to:
\begin{equation}
\label{uncert0}
\left| l_{\rm Pl} \phi_i  \right| \ge  \frac{2 \times 10^6}{j^{15/2} \left|
l_{\rm Pl}^2  \dot{\phi}_i \right| }  .
\end{equation}

The inflaton is effectively localized around the minimum of the
potential within the range $| \Delta \phi |$, and a given initial
condition can be set anywhere within this range,
$\phi_i \approx \pm \Delta \phi$. The sign of the field's kinetic energy
at the beginning of the semi--classical regime is important. If ${\rm
sgn} ( l_{\rm Pl}^2\dot{\phi}_i) = +1$, the field 
begins moving up its potential immediately. If, on the other 
hand, ${\rm sgn} (  l_{\rm Pl}^2 \dot{\phi}_i ) =-1$, 
the inflaton rolls back rapidly through the minimum of its potential
and up the other side. Consequently, there are 
two separate possibilities if the potential is an even function of the 
field: 
$\{ \phi_i \approx \left| \Delta \phi \right|, \dot{\phi}_i >0 \}$ or
$\{ \phi_i \approx - \left| \Delta \phi \right|, \dot{\phi}_i
>0 \}$. In the following Sections we concentrate on these initial conditions
and also consider the set $\{ \phi_i =0 , \dot{\phi}_i >0 \}$, as this represents 
the midway point between the two extremes.

\section{Analytical Approximation Scheme}

\subsection{Transition from Semi--Classical to Classical Dynamics}

In this Section, we develop an approximate analytical 
approach to estimating the conditions for successful inflation in LQC
when the field is initially located in the vicinity of 
its potential minimum. Both the {\sc Ham} and {\sc Fried} 
quantization schemes are considered, where the scalar field 
equation of motion is given by Eq. (\ref{scalareom}). 
The basis of the approximation is to separate the rolling of the field 
to its maximal value $\phi_{\rm max}$ into 
two distinct epochs -- a semi--classical, super--inflationary phase
followed by a classical epoch. The asymptotic form (\ref{approxD})
of the eigenvalue function, $D_l (a)$, is invoked throughout the 
semi--classical era and it is assumed that the transition 
to classical dynamics occurs instantaneously when
$D_l$ reaches unity. It is further assumed that once this condition 
has been attained, $D_l$ remains fixed at unity. Numerical 
simulations confirm this by indicating 
that once the eigenvalue function approaches unity 
it does so very rapidly.

The field reaches its point of maximal displacement when the 
potential begins to dominate its kinetic energy. 
Numerical solutions indicate that a good estimate for the turning 
point can be determined from the condition \cite{madsencoles} 
\begin{equation}
\label{fieldmax}
\frac{1}{2} \dot{\phi}^2_{\rm max} \approx  V \left( \phi_{\rm max} 
\right)  .
\end{equation}
For concreteness, we consider a 
quadratic self--interaction potential, $V=m^2\phi^2/2$, 
where $m$ represents the mass of the field, 
although the approach we develop 
is independent of the particular functional form 
of the inflaton potential. A quadratic potential 
may also be viewed as a lowest--order
Taylor expansion of a more general potential around a turning point.
Moreover, since the inflaton is evolving away 
from the minimum, it is expected that 
its kinetic energy will dominate the cosmic dynamics 
until Eq. (\ref{fieldmax}) applies. We therefore view 
the inflaton as a massless field $(V=0)$ until it reaches its turning 
point.  

{}From a phenomenological point of view, there are two important 
constraints that must be satisfied for successful inflation. 
Firstly, sufficient inflation must occur to solve the 
horizon problem and this implies that the field must be sufficiently 
displaced from its minimum when it begins to roll back down. 
The required amount of inflation is dependent on the reheating temperature, 
although $60$ e--foldings is typically required. (We do not 
take into account the e-foldings of accelerated expansion that arise 
during the superinflationary epoch. This is because standard perturbation
theory is unstable during this phase \cite{tsm03}. 
We therefore implicitly assume that 
anisotropies in the CMB are generated during a conventional phase of
slow--roll inflation).  
For a quadratic potential, the COBE normalization 
of the CMB power spectrum constrains the
inflaton mass to be $m \approx 10^{-6} l_{\rm Pl}^{-1}$
and, in this case, the horizon problem is solved if 
$l_{\rm Pl} \phi_{\rm max} \ge 3$ \cite{linde,lidlyth}.
 
The second constraint concerns 
the region of parameter space where the semi--classical and 
classical approximations are valid. In this work, we confine 
ourselves to the regime where  
the dynamics is determined through coupled ODEs
and spacetime is effectively viewed as a continuum. Consequently, 
at the epoch of transition to the classical regime,
the Hubble length should exceed the limiting scale 
consistent with such an approximation, i.e., 
$H^{-1}> \sqrt{\gamma} l_{\rm Pl}$. (This is equivalent to 
the condition that a classical description 
of the dynamics is only consistent at energy scales below the 
Planck scale.) Since the Hubble parameter and the inflaton's 
kinetic energy are  monotonically increasing functions during the 
semi--classical regime, this leads to an upper bound on the 
duration of that phase.  
An estimate for the limit on the field's kinetic energy at the transition 
epoch follows directly from the Friedmann equation by setting 
$D_l =1$. For both the {\sc Ham} and {\sc Fried} quantization schemes, 
this implies that $H_S^2 \approx 4 \pi l_{\rm Pl}^2 \dot{\phi}_S^2/3$ and 
hence that 
\begin{equation}
\label{genupper}
\left| l_{\rm Pl}^2 \dot{\phi}_S \right| \le \left( \frac{3}{4\pi \gamma}
\right)^{1/2}   ,
\end{equation}
where a subscript $S$ denotes values of the parameters 
at the transition time. 
We refer to the bound (\ref{genupper}) as the {\em kinetic bound}.

\subsection{Classical Dynamics}

To proceed, let us consider the classical phase. 
For a massless scalar field $(V=0)$, the Friedmann and scalar field 
equations 
can be expressed in the Hamilton--Jacobi form: 
\begin{eqnarray}
\label{HJ1}
\left( \frac{dH}{d\phi} \right)^2 = 12 \pi l_{\rm Pl}^2 H^2
\\
\label{HJ2}
\frac{dH}{d\phi} = - 4 \pi l_{\rm Pl}^2 \dot{\phi}  ,
\end{eqnarray}
where time derivatives are replaced throughout by derivatives 
with respect to the scalar field.  
The general solution to Eqs. (\ref{HJ1}) and (\ref{HJ2}) is given by 
\begin{equation}
\label{genHsol}
H=H_S \exp \left[ - \sqrt{12\pi} l_{\rm Pl} 
\left( \phi -\phi_S \right) \right]  .
\end{equation}
Substituting  Eqs. (\ref{HJ2}) and (\ref{genHsol}) into Eq. (\ref{fieldmax})
implies that the maximal value attained by the field 
for both {\sc Ham} and {\sc Fried} schemes is given by 
\begin{equation}
\label{phiSmax}
\phi_{\rm max} e^{\sqrt{12 \pi} l_{\rm Pl} \phi_{\rm max}}
 \approx \frac{\left| \dot{\phi}_S \right|}{m}
e^{\sqrt{12 \pi} l_{\rm Pl} \phi_S} .
\end{equation}
When the potential is negligible,
the scalar field equation (\ref{scalareom}) 
admits the first integral: 
\begin{equation}
\label{kineticsol}
\dot{\phi} = \dot{\phi}_i \left( \frac{a}{a_i} \right)^{3/(1-l)} ,
\end{equation}
where a subscript $i$ denotes initial values at the beginning of the 
semi--classical regime. Substituting Eq. (\ref{kineticsol}) 
into Eq. (\ref{phiSmax}) then implies that
\begin{equation}
\label{genphimax}
\phi_{\rm max} e^{\sqrt{12 \pi} l_{\rm Pl} \phi_{\rm max}} 
\approx \frac{\left| \dot{\phi}_i \right|}{m} 
\left( \frac{a_S}{a_i} \right)^{3/(1-l)}
e^{\sqrt{12 \pi}l_{\rm Pl} \phi_S}  .
\end{equation}

As discussed in \S 2.1, the smallest value for the scale factor 
that is
consistent with viewing spacetime as a continuum is  
\begin{equation}
\label{initialscale}
a_i \approx \sqrt{\gamma} l_{\rm Pl} , \qquad 
\frac{a_i}{a_*} \approx \left( \frac{3}{j} \right)^{1/2}
\end{equation}
and it then follows from Eq. (\ref{approxD}) that 
\begin{eqnarray}
\label{approxaS}
\frac{a_S}{a_*} \approx \left( \frac{l+1}{3} \right)^{1/(4-2l)} 
\nonumber \\
\frac{a_S}{a_i} \approx \left( \frac{j}{3} \right)^{1/2} 
\left( \frac{l+1}{3} \right)^{1/(4-2l)}  .
\end{eqnarray}
%\frac{a_S}{a_*} \approx \left( \frac{7}{12} \right)^{6/15} 
%\approx 0.8 , 
%\qquad \frac{a_S}{a_i} \approx 0.46 j^{1/2}
Substituting Eq. (\ref{approxaS}) into (\ref{kineticsol}) 
then yields an estimate for the initial value of the 
field's kinetic energy in terms of its value 
at the transition time:  
\begin{equation}
\dot{\phi}_i = \dot{\phi}_S \left( \frac{3}{j} \right)^{3/(2-2l)} 
\left( \frac{3}{l+1} \right)^{3/[(4-2l)(1-l)]}   .
\end{equation}

Imposing the kinetic bound (\ref{genupper}) then 
leads to an estimate for an upper limit on 
the combination of parameters $j | l_{\rm Pl}^2 \dot{\phi}_i 
|^{2(1-l)/3}$ in terms of a constant with numerical value determined 
in terms of the parameters $\gamma$ and $l$:
\begin{equation}
\label{generalkinetic}
j \left| l_{\rm Pl}^2 \dot{\phi}_i \right|^{2(1-l)/3} \le 
3 \left( \frac{3}{4\pi \gamma} \right)^{(1-l)/3} 
\left( \frac{3}{1+l} \right)^{1/(2-l)}   .
\end{equation}
Eq. (\ref{generalkinetic}) simplifies to 
\begin{equation}
\label{kineticlimit}
j \le \frac{5}{\left| l_{\rm Pl}^2 \dot{\phi}_i \right|^{1/6}}
\end{equation}
for $l= 3/4$ and to 
\begin{equation}
\label{kineticlimitsmall}
j \le \frac{6.4}{\left| l_{\rm Pl}^2 \dot{\phi}_i \right|^{2/3}} 
\end{equation}
for $l \ll 1$.

We now require the value of the scalar field at the transition epoch 
in order to estimate the maximal value of the scalar field (\ref{genphimax}) 
in terms of its initial value.  
This is determined from the solution to the field equations 
for each of the quantization schemes. 

\subsection{{\sc Ham} Quantization} 

The solution to the Friedmann equation (\ref{fredeq1a}), neglecting the 
potential, is
\begin{equation}
\label{genlsolution}
\phi =\phi_i +B_l \left[ \left( \frac{a}{a_i} 
\right)^{3(2-l)/(2-2l)} -1 \right]   ,
\end{equation}
where
\begin{eqnarray}
\label{genlB}
B_l &=& \frac{2(1-l)}{3(2-l)} \left( \frac{6}{8\pi l_{\rm Pl}^2} 
\right)^{1/2} 
\nonumber \\
&&\times \left( \frac{3}{l+1} \right)^{3/(4-4l)}
\left( \frac{a_i}{a_*} \right)^{3(2-l)/(2-2l)}
\end{eqnarray}
is a constant. 
Since the expressions for general $l$ are cumbersome, 
we focus in what follows on the value $l = 3/4$ and the 
limit $l \ll 1$. We discuss the limit $l \rightarrow 1$ in \S
6. Eq. (\ref{genlsolution}) simplifies to 
\begin{eqnarray}
\label{Bojosol}
\phi =\phi_i +B \left[ \left( \frac{a}{a_i} \right)^{15/2} -1 
\right] \\
\label{defB}
B_{3/4} \equiv \frac{2}{15} \left( \frac{6}{8\pi l_{\rm Pl}^2} \right)^{1/2}
\left( \frac{12}{7} \right)^3 \left( \frac{a_i}{a_*} \right)^{15/2}
\end{eqnarray}
for $l=3/4$ and to 
\begin{eqnarray}
\label{phi0}
\phi =\phi_i +B_0 \left[ \left( \frac{a}{a_i} \right)^3 -1 \right] \\
B_0 = 3^{-1/4} \left( \frac{6}{8\pi l_{\rm Pl}^2} \right)^{1/2} 
\left( \frac{a_i}{a_*} \right)^3
\end{eqnarray}
for $l \ll 1$.

In general, the total shift in the value of the field 
induced by the anti--frictional effect of the semi--classical 
phase increases for increasing $j$, since the duration of 
the super--inflationary 
dynamics is enhanced for higher values of $j$. Consequently, 
the condition for the horizon problem to be solved can be expressed 
as a lower limit on the value of $j$ for 
given values of $\{ l, \dot{\phi}_i \}$.  

Substituting Eqs. (\ref{Bojosol}) and (\ref{defB}) 
into Eq. (\ref{genphimax}) and employing the estimates 
(\ref{initialscale}) and (\ref{approxaS}) implies that 
\begin{equation}
\label{Bphimax}
l_{\rm Pl} \phi_{\rm max} e^{\sqrt{12 \pi} l_{\rm Pl} \phi_{\rm max}}
\approx 140 j^6 \left| l_{\rm Pl}^2 
\dot{\phi}_i \right| e^{\sqrt{12 \pi} l_{\rm Pl} \phi_i}
\end{equation}
when $l=3/4$, 
where it is assumed that $j$ is sufficiently large for $(a_S/ a_i)^{15/2}
\gg 1$ (this requires $j \ge {\cal{O}}(3)$). The COBE normalization 
constraint on the mass of the inflaton field has also been imposed.  
The horizon problem is therefore solved ($l_{\rm Pl} \phi_{\rm max} \ge 3$)
if 
\begin{equation}
\label{HPphiinitial}
\ln \left( j^6 \left| l_{\rm Pl}^2 \dot{\phi}_i \right| \right) + 
\sqrt{12 \pi} l_{\rm Pl} \phi_i \ge 14.6  .
\end{equation}

As discussed in \S 3, an estimate for the initial value of the 
inflaton field may be derived from the uncertainty principle. 
Since we are primarily interested in this Section in determining 
the influence of the anti--frictional effects, we specify $\phi_i =0$
as this provides a transparent measure of 
the overall shift in the value of the field due to the semi--classical 
corrections. In effect, the uncertainty principle then 
changes this value by a constant amount for 
each set of $\{ j, l, \dot{\phi}_i \}$. The combined effects of the
anti--friction and uncertainty principle on the constraints are 
determined by numerical analysis in following Section.

Eq. (\ref{HPphiinitial}) then implies that
\begin{equation}
\label{Bhorizon}
j \ge \frac{11}{\left| l_{\rm Pl}^2 \dot{\phi}_i \right|^{1/6}}
\end{equation}
for $\phi_i =0$ and comparing the limits (\ref{Bhorizon}) and
(\ref{kineticlimit}) for the {\sc Ham} quantization implies that they
are incompatible for this value of $l$. This would seem to indicate
that successful inflation within a purely semiclassical description
is not possible with these initial conditions.

It is worth addressing briefly
the question of how different values of the parameter
$l$ would alter this conclusion. Since lowering the value
of $l$ leads to superinflationary expansion that is closer to the
exponential limit, it might be expected that the kinetic energy of the
inflaton field would grow less rapidly during the semi--classical phase.
However, Eq. (\ref{kineticlimitsmall})
implies that lowering $l$ does not significantly
weaken the kinetic bound on $j$. Furthermore, for $l \ll 1$, substituting
Eq. (\ref{phi0}) into Eq. (\ref{genphimax}) implies that the
horizon problem is only solved if
\begin{equation}
\label{only}
\ln \left( j^{3/2} \left| l_{\rm Pl}^2 \dot{\phi}_i \right| \right)
+ \sqrt{12 \pi} l_{\rm Pl} \phi_i  \ge 7.2
\end{equation}
and for $\phi_i =0$, the constraint (\ref{only})
reduces to the condition
\begin{equation}
\label{small83}
j \ge \frac{120}{\left| l_{\rm Pl}^2 \dot{\phi}_i \right|^{2/3}}  .
\end{equation}

\subsection{{\sc Fried} Quantization}

For the {\sc Fried} quantization scheme,
the solution to the Friedmann equation (\ref{fredeq2a})
in the limit of kinetic energy domination is
\begin{equation}
\label{Hsol}
\phi =\phi_i + \left( \frac{3}{4 \pi l_{\rm Pl}^2} \right)^{1/2} \ln
\left( \frac{a}{a_i} \right)
\end{equation}
and substituting Eq. (\ref{Hsol}) into Eq. (\ref{genphimax}) implies
that
\begin{equation}
\label{Hphimax}
\phi_{\rm max} e^{\sqrt{12 \pi} l_{\rm Pl} \phi_{\rm max}}
\approx \frac{\left| \dot{\phi}_i\right|}{m} \left( \frac{a_S}{a_i}
\right)^{3(2-l)/(1-l)}
e^{\sqrt{12 \pi} l_{\rm Pl} \phi_i}  .
\end{equation}

The method of estimating when the horizon problem is solved
is similar to that employed in \S 4.3 for {\sc Ham}
quantization. Substituting Eqs. (\ref{initialscale}) and (\ref{approxaS})
into Eq. (\ref{Hphimax}) and imposing the requirement that $l_{\rm Pl}
\phi_{\rm max} \ge 3$ implies that
\begin{eqnarray}
\label{3/4max}
\ln \left( j^{15/2} \left| l_{\rm Pl}^2 \dot{\phi}_i \right| \right)
+ \sqrt{12\pi} l_{\rm Pl} \phi_i \ge 17.2 , \qquad l=\frac{3}{4}
\\
\ln \left( j^3 \left| l_{\rm Pl}^2 \dot{\phi}_i \right| \right)
+ \sqrt{12\pi} l_{\rm Pl} \phi_i \ge 10.6 , \qquad l \ll 1 .
\end{eqnarray}

For the case where $\phi_i =0$,
this implies that the horizon problem is solved if
\begin{eqnarray}
\label{Hhorizon}
j \ge \frac{10}{\left| l_{\rm Pl}^2 \dot{\phi}_i \right|^{2/15}} ,
\qquad l=\frac{3}{4} \\
\label{Hhorizonsmall}
j \ge \frac{35}{\left| l_{\rm Pl}^2 \dot{\phi} \right|^{1/3}} ,
\qquad l \ll 1  .
\end{eqnarray}

Comparing the limits (\ref{Hhorizon}) and (\ref{Hhorizonsmall})
with the corresponding kinetic bounds (\ref{kineticlimit})
and (\ref{kineticlimitsmall}) implies that
\begin{eqnarray}
\label{compare}
\frac{10}{\left| l_{\rm Pl}^2 \dot{\phi}_i \right|^{2/15}} \le j \le
\frac{5}{\left| l_{\rm Pl}^2 \dot{\phi}_i \right|^{1/6}} , \qquad l=
\frac{3}{4} \\
\label{powersmall}
\frac{35}{\left| l_{\rm Pl}^2 \dot{\phi}_i \right|^{1/3}} \le j \le
\frac{6.4}{\left| l_{\rm Pl}^2 \dot{\phi}_i \right|^{2/3}} , \qquad l \ll 1 .
\end{eqnarray}
As a result, the horizon problem can only
be solved if $\left| l_{\rm Pl}^2 \dot{\phi} \right| \le 10^{-9}$
when $l=3/4$ and if $\left| l_{\rm Pl}^2 \dot{\phi} \right| \le
6 \times 10^{-3}$ when $l \ll 1$.

%-----------------------------
\section{Numerical Results}
%-------------------------------

In this Section, we determine the regions of parameter space
that lead to successful inflation in
the {\sc Ham} and {\sc Fried} quantization schemes by
numerically integrating the field equations, where the complete
expression (\ref{defD}) is assumed for the eigenvalue function
and the inflaton potential is included.
The results are presented in the form of plots of the ambiguity parameter,
$j$, against $\dot{\phi_{i}}$ for a given value of $l$.
On each plot a solid line represents the boundary for
the horizon problem to be just solved (and consequently
for large angular scales on the CMB to correspond to the
turning point in the field's dynamics). 
A dashed line represents the boundary where the
kinetic bound is just violated. Shaded areas represent 
regions for successful inflation.

\subsection{$ \{ \phi_i =0, \dot{\phi}_i >0 , l = 3/4 \}$}

We begin by considering the set of initial conditions
$\{ \phi_i =0, \dot{\phi}_i >0 \}$ with $l=3/4$ in order to compare the 
exact numerical results with the approximation scheme developed 
in \S 4. 
Fig. (\ref{figure2.eps}) shows the
analytic estimates for {\sc Ham} quantization 
without the minimum uncertainty (\ref{uncert0}) imposed 
and Fig. (\ref{figure3.eps}) shows the
results for the same system from numerical integration.
The corresponding results for 
{\sc Fried} quantization are shown in Figs. 
(\ref{figure4.eps}) and (\ref{figure5.eps}), respectively.
The horizon problem is solved 
above the solid line and the kinetic constraint is satisfied
below the dashed line. Necessary conditions for successful
{\sc Fried} inflation are $| l_{\rm Pl}^2 
\dot{\phi}_i | \leq 10^{-8}$ and 
$j \sim 100$.

There is good agreement between the analytic and numerical approaches
in both schemes. The analytic approximation typically underestimates
the maximum value of the scalar field by about $0.1 l_{\rm Pl}^{-1}$
when $\phi_{\rm max} \approx 3 l_{\rm Pl}^{-1}$ leading to a small
error in the total number of e--foldings, $\Delta N \approx 4$. Such
an error is comfortably within other uncertainties that reduce the
total number of e--foldings required to solve the horizon problem. In
particular, the required number of e--foldings may be as low as $N
\approx 30$  for a reheating
temperature at the electroweak scale. The analytic approximation works
well because the transition from $D_l
\propto a^{3(2-l)/(1-l)}$ to $D_l \rightarrow 1$ is very
rapid. Numerical results also indicate that the acceleration of the
scalar field is initially very strong and that the non--inflationary
phase where the field rolls slowly up the potential is relatively
short.

\subsection{$\{ \phi_i \ne 0, \dot{\phi}_i >0 , l = 3/4 \}$}

We now discuss the consequences of the ambiguities for realizing successful 
inflation from the initial conditions imposed by the 
uncertainty principle (\ref{uncert0}). 
For the {\sc Ham} quantization scheme, substituting 
Eq. (\ref{uncert0}) into the condition 
for the horizon problem to be solved, Eq. (\ref{HPphiinitial}), implies that 
the condition 
\begin{equation}
\label{Buncertinit}
\ln \left| j^6 l_{\rm Pl}^2 \dot{\phi}_i \right|
\pm \frac{1.2 \times 10^7}{j^{15/2} \left| l_{\rm Pl}^2 \dot{\phi}_i
\right|} > 14.6
\end{equation}
leads to successful inflation.
The effect of starting the dynamics away from the minimum is contained
within the second term on the left--hand side. Thus,
for a given initial kinetic energy, the horizon problem
is solved for sufficiently small $j$.
The corresponding condition for {\sc Fried} quantization
follows by substituting Eq. (\ref{uncert0}) into Eq. (\ref{3/4max}):
\begin{equation}
\label{Huncertinit}
2.5 \ln \left( \frac{j^{15/2} \left| l_{\rm Pl}^2 \dot{\phi}_i
\right|}{3.2\times
10^7} \right) \pm \frac{3.2 \times 10^7}{j^{15/2}\left| l_{\rm Pl}^2
\dot{\phi}_i \right|} >0  .
\end{equation}

Fig. (\ref{figure6.eps}) shows the numerical results
for the {\sc Ham} quantization scheme,
where the inflaton field emerges into the semi--classical 
regime such that $\{ \phi_i , \dot{\phi}_i \} >0$. 
In this case,  there are two pairs of constraints. 
The regions above the solid line and
below the dotted line solve the horizon problem.
The upper region solves the horizon problem due to 
$j$ and $\dot{\phi}_i$ being sufficiently large. The 
lower region solves the horizon problem since  
the inflaton's starting position is higher up the potential
for smaller $j$.  
The region above the dashed line violates the kinetic
bound. At small $j$, there is a further constraint that must be considered. 
In this region of parameter space, 
the field may be displaced from its minimum to such an extent 
that most of its energy is in the form of its potential. 
In this case, the potential is bounded to be less than 
the Planck scale. 
The region below the dot-dashed line violates this constraint.
Thus, there exists a region for successful inflation 
in {\sc Ham} quantization.

Fig. (\ref{figure7.eps}) illustrates the
numerical results for {\sc Fried} quantization. 
The horizon problem is solved for all values of $\{ j , \dot{\phi}_i \}$
covered in the figure. The kinetic bound and corresponding 
constraint on the potential limit the region of parameter space. 
Successful inflation is possible.  

\subsection{Effects of varying $l$}

We now consider how different values of $l$ alter the above
conclusions.  We have numerically integrated the field equations where
$l$ varies in the range $0.01 \le l \le 0.95$.  (The superinflationary
expansion is so extreme for higher $l$ that the numerical integration
becomes unreliable).  In \S 4.4, it was found that in the case of {\sc
Fried} quantization, the region of parameter space for successful
inflation is widened for smaller values of $l$.  This follows 
because the power dependences on the initial kinetic energy in
Eq. (\ref{powersmall}) differ to a greater degree as $l$
decreases and the intersection of the two constraints
in Fig. (\ref{figure4.eps}) is located at higher values of $\{ j,
\dot{\phi}_i \}$. A lower value of $l$ corresponds
to an expansion rate that is closer to the exponential limit and
therefore the kinetic energy of the field grows less
rapidly. Consequently, the superinflation phase must last longer to
ensure the field has sufficient kinetic energy to solve the horizon
problem.  The full numerical integration supports this generic
behavior. The agreement between the analytic and numerical approaches
improves at higher $l$, which can be seen from Fig.\ (\ref{figure8.eps})
showing that the peak widens for small $l$.  At lower $l$, the turning
point of the field is underestimated by no more than $0.1 l_{\rm
Pl}^{-1}$ to $0.2 l_{\rm Pl}^{-1}$ when $\phi_{\rm max} \approx 3
l_{\rm Pl}^{-1}$.

For the {\sc Ham} quantization scheme, the kinetic bound
and the horizon problem constraint both take the
form $j | l_{\rm Pl}^2 \dot{\phi}_i |^{2(1-l)/3}
\approx A_k$, where $A_k = A_k (l) $ is a numerical constant determined
by $l$. In this case, it is the numerical factor
$A_k$
which is important and successful inflation
requires $A_{\rm kinetic} >
A_{\rm horizon}$.
The analytic approach for the case of $\phi_i = 0$, as summarized in Eqs.
(\ref{kineticlimitsmall}) and (\ref{small83}),
indicates that reducing $l$ below $l=3/4$ strengthens the inequality
$A_{\rm kinetic} < A_{\rm horizon}$.
For $l> 3/4$, numerical integration
implies that the difference in the numerical factors is reduced,
but not sufficiently for the inequality to be reversed, at least
up to $l \approx 0.95$.
Hence, the two lines never intersect in Fig. ({\ref{figure3.eps})
and there is no region of parameter space which simultaneously
satisfies both bounds. Indeed, for given values of $\{ \dot{\phi}_i ,l \}$, 
the highest value of $j$ that is just consistent with 
the kinetic bound typically leads to a turning point  
in the field's motion at 
$\phi_{\rm max} \approx 2.4 l_{\rm Pl}^{-1}$. Numerical integration 
indicates that 
this holds over a wide range of $l$ and implies that the field 
must be displaced from its minimum for successful inflation to proceed. 

As in {\sc Fried} quantization,  the agreement
between analytic and numerical results is good, and
improves at higher $l$. 
We conclude, therefore, that varying $l$
does not significantly alter the overall qualitative 
picture in this scenario.

%____________________________________________________________________
\begin{figure}
\includegraphics[width=8.5cm]{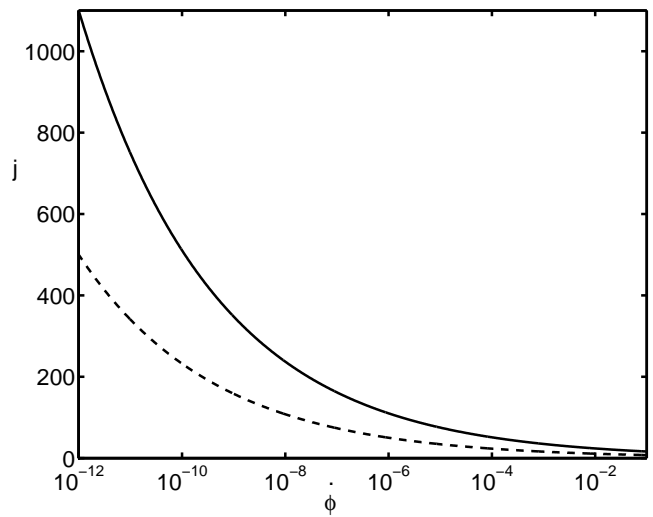}%
%\centerline{\def\epsfsize#1#2{0.6#1}\epsffile{BnUb.eps}}
\caption[] {\label{figure2.eps}
Analytic results for {\sc Ham} quantization with $l=3/4$
and initial conditions $\phi_i =0$ and $\dot{\phi}_i >0$. 
The solid line corresponds to the case where the
turning point of the inflaton is at $\phi = 3 l_{\rm Pl}$. 
Sufficient inflation to solve the horizon problem 
arises in the region above this line. 
The kinetic bound is satisfied in the region 
below the dashed line. The two regions do not overlap.}
\end{figure}
%_____________________________________________________________________

%____________________________________________________________________
\begin{figure}
\includegraphics[width=8.5cm]{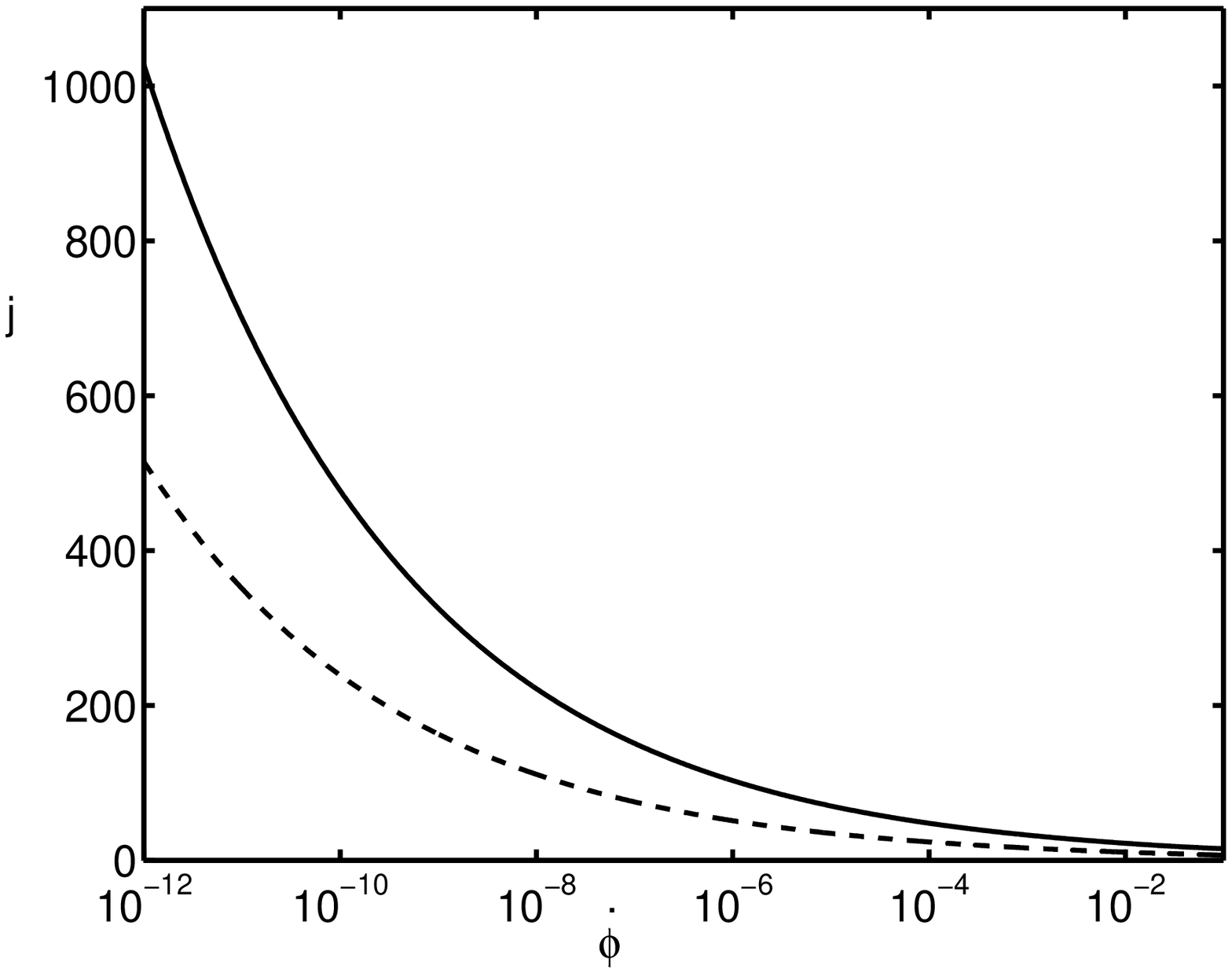}%
%\centerline{\def\epsfsize#1#2{0.6#1}\epsffile{HnUb.eps}}
\caption[] {\label{figure3.eps}
Numerical results corresponding to the scheme 
shown in Fig. \ref{figure2.eps}.}
\end{figure}
%_____________________________________________________________________

%____________________________________________________________________
\begin{figure}
\includegraphics[width=8.5cm]{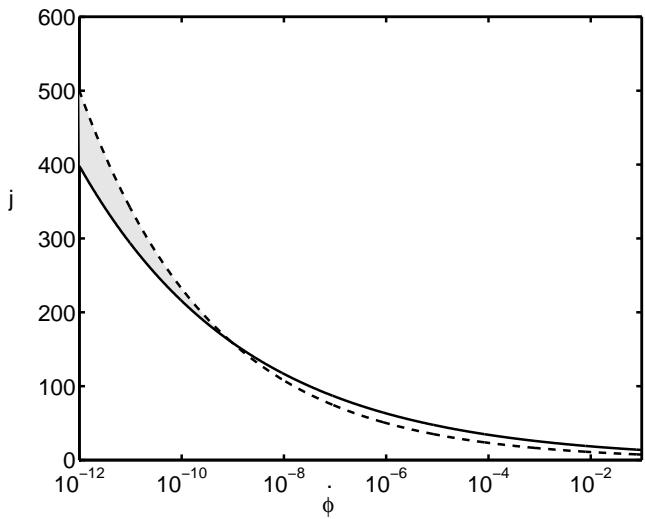}%
%\centerline{\def\epsfsize#1#2{0.6#1}\epsffile{HnUb.eps}}
\caption[] {\label{figure4.eps}
Analytic results for {\sc Fried} quantization with $l=3/4$ and  
initial conditions $\phi_i =0$ and $\dot{\phi}_i >0$. 
The horizon problem is solved in the region 
above the solid line and the kinetic bound is satisfied
below the dashed line. The shaded area
represents the region of parameter space that leads to successful inflation 
for this set of initial conditions.}
\end{figure}
%_____________________________________________________________________

%____________________________________________________________________
\begin{figure}
\includegraphics[width=8.5cm]{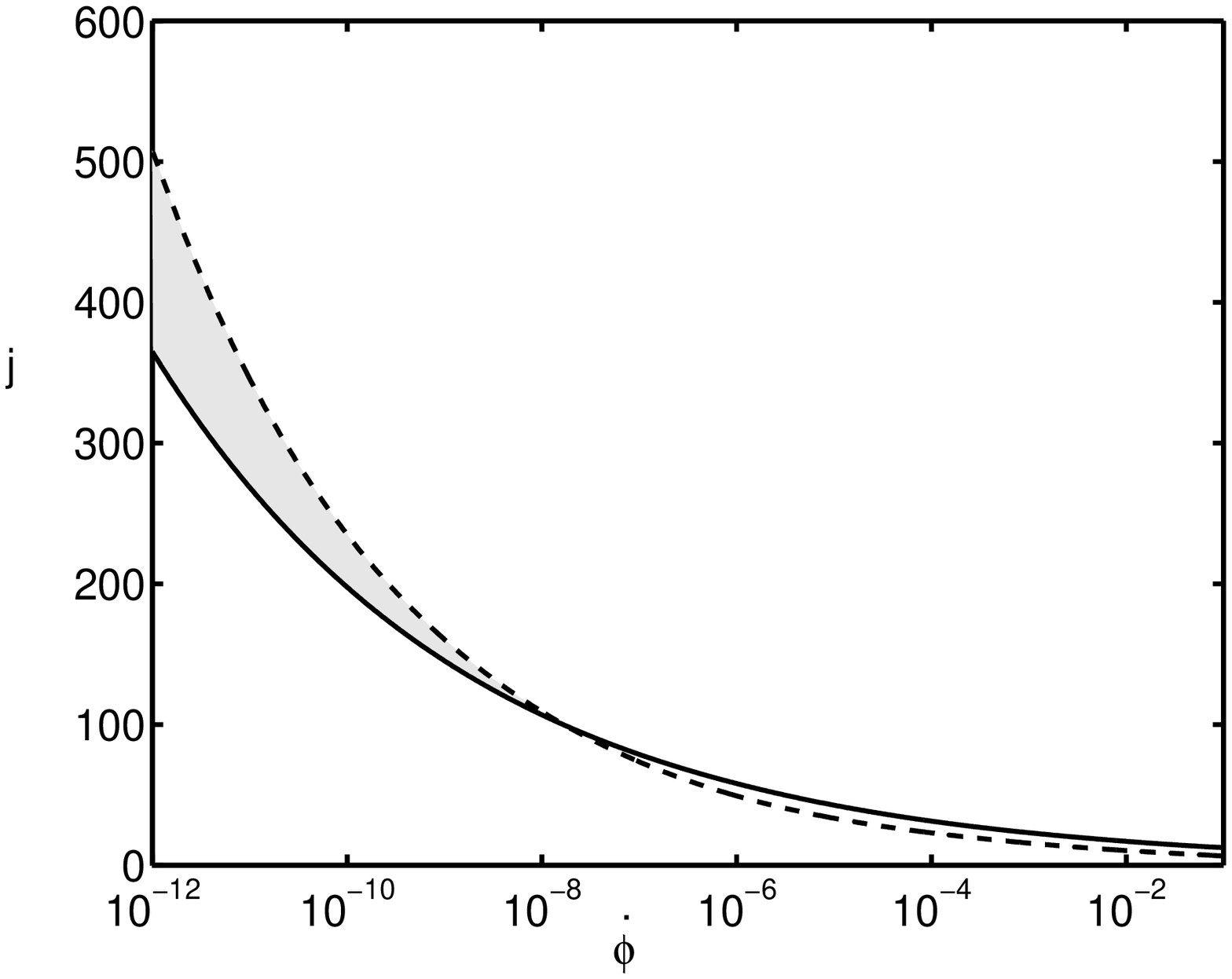}%
%\centerline{\def\epsfsize#1#2{0.6#1}\epsffile{HnUa.eps}}
\caption[] {\label{figure5.eps}
Numerical results corresponding to the scheme shown in
Fig \ref{figure4.eps}.}
\end{figure}
%_____________________________________________________________________

%_____________________________________________________________________
\begin{figure}
\includegraphics[width=8.5cm]{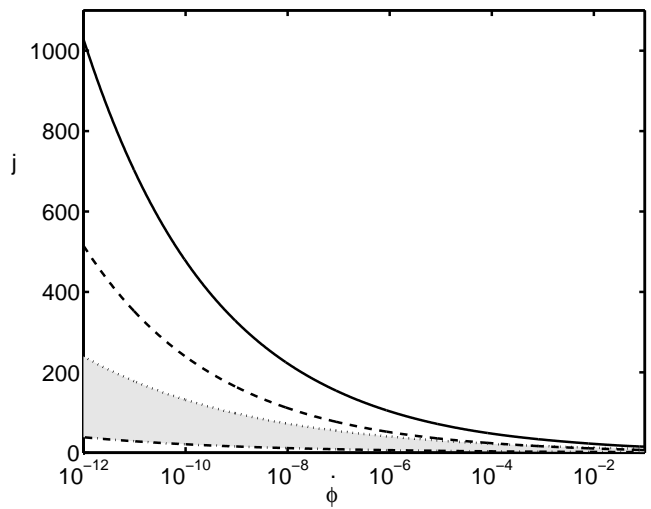}%
%\centerline{\def\epsfsize#1#2{0.6#1}\epsffile{BwUa.eps}}
\caption[] {\label{figure6.eps}
Numerical results for {\sc Ham} quantization with $l=3/4$, 
where the initial value of the inflaton is determined by imposing the 
uncertainty principle (\ref{uncert0}). 
The horizon problem is solved 
in regions above the solid line and below the dotted line.
The region below the dashed line satisfies the kinetic bound.
The potential energy of the field exceeds the Planck scale
below the dot-dashed line. Successful inflation is
possible in the shaded region.}
\end{figure}
%_____________________________________________________________________

%_____________________________________________________________________

\begin{figure}
\includegraphics[width=8.5cm]{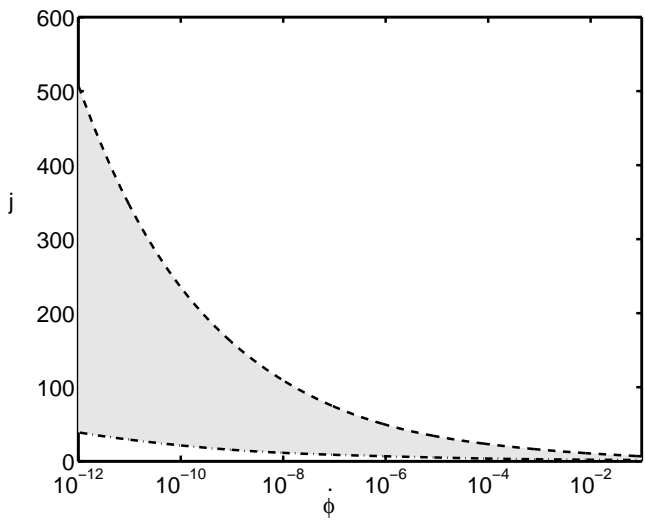}%
%\centerline{\def\epsfsize#1#2{0.6#1}\epsffile{HwUa.eps}}
\caption[] {\label{figure7.eps}
Numerical results for {\sc Fried} quantization with $l=3/4$,
where the initial value of the inflaton is determined by imposing the 
uncertainty principle (\ref{uncert0}). 
The horizon problem is solved for 
all values of $j$ and $\dot{\phi}_i$ covered in the figure.
The conditions for successful inflation are therefore limited only 
by the requirement that the classical phase corresponds to energy scales 
below the Planck scale. Successful inflation arises in the shaded region.}
\end{figure}
%_____________________________________________________________________

\begin{figure}
\includegraphics[width=8.5cm]{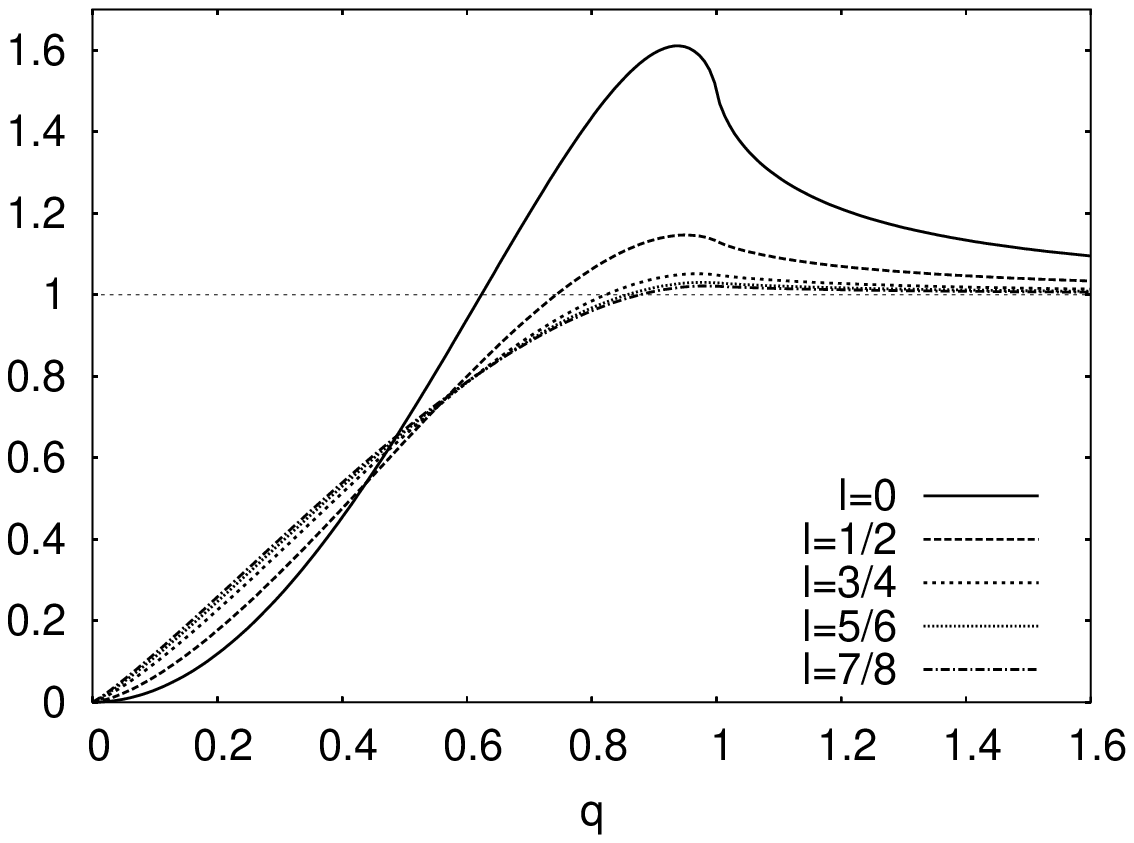}%
\caption{\label{figure8.eps}
The behavior of the function $D_l$ for different values of $l$.}
\end{figure}

\section{Discussion}

In this paper we have considered some of the cosmological consequences of 
ambiguities that arise in loop quantum gravity. We have focused 
primarily on the importance of these ambiguities in realizing the conditions 
that lead to a phase of slow--roll inflation when the inflaton is 
initially located in a minimum of its potential. We have invoked a
semi--classical treatment, where the dynamics of the universe is 
governed by non--linear ODEs. The key requirement 
for successful inflation within this framework 
is that sufficient inflation is possible in the region of 
parameter space where this approximation remains valid. In particular, 
this implies that the initial kinetic energy of the inflaton 
must be sufficiently small. 

Our main conclusion is that the initial conditions for slow--roll
inflation can be realized in a wide region of parameter space in loop
quantum cosmology. In this sense, therefore, LQC is
robust to ambiguities in the quantization and enhances the allowed
range of initial conditions for inflation. In particular, kinetic
energies many orders of magnitude below the Planck scale can lead to
inflation. Moreover, parameters can be chosen such that the turning
point of the inflaton is near to $ 3 l_{\rm Pl}^{-1}$, corresponding
to the largest scales accessible to observations. Equation
(\ref{Bphimax}) shows that $\phi_{\max}$ depends only logarithmically
on $j$ and $\dot{\phi}_i$ and thus is sensitive only to the order of
magnitude of those values. (Indeed, by using the Lambert function $W(x)$
defined such that $W(x)e^{W(x)}=x$ and behaving like a logarithm for $x>1$,
we have $l_{\rm Pl}\phi_{\rm max}=W(140\sqrt{12\pi} j^6| \dot{\phi}_i
l_{\rm
Pl}^2|e^{\sqrt{12\pi}l_{\rm Pl}\phi_i})$.) This verifies what can
already be observed from Fig.\ 2 of Ref. \cite{tsm03}.

For the choice $l=3/4$, conditions for successful inflation can be
achieved in the {\sc Fried} quantization scheme if $j$ is sufficiently
high even when the field is located at the potential minimum. The
field tends to move further up the potential in the {\sc Fried}
quantization than in {\sc Ham} quantization. In this sense, {\sc
Fried} quantization might be favored from a phenomenological point of
view at the semi--classical level, as it results in a larger region of
parameter space for successful inflation (this has already been
indicated in \cite{golam}). This is interesting given that the {\sc
Ham} quantization scheme is directly based on a Hamiltonian whereas
{\sc Fried} involves additional multiplication with eigenvalues of the
geometrical density operator and thus appears less natural conceptually.

In {\sc Ham} quantization, the field needs to be displaced 
from its minimum, for example by quantum uncertainty effects,
if it is to move sufficiently 
far up the potential without violating the bounds imposed 
by the semi--classical approximation. 
However, care should be taken in interpreting the kinetic bound 
(\ref{kineticlimit}) -- 
the solution (\ref{kineticsol}) overestimates the value of the 
field's kinetic energy at the transition since the form 
of the eigenvalue function, $D_l$, given 
in Eq. (\ref{approxD}) represents its asymptotic form in the limit 
$a \ll a_*$. Moreover, the estimate for $a_S$, 
the scale factor at the transition epoch, given in Eq. (\ref{approxaS}),
is not precise, since there is no exact definition 
for this parameter. If the kinetic bound 
(\ref{kineticlimit}) were to be relaxed slightly, then
it would become marginally consistent with the 
horizon problem constraint 
(\ref{Bhorizon}). This indicates that {\em if} successful inflation could be 
realized within this semi--classical framework,
the conditions would be such that just enough e--folds 
of accelerated expansion would arise for 
the horizon problem to be solved. As a result, astrophysically observable 
scales today would 
correspond to the turning point in the field's dynamics. 

It should be emphasized that failure to satisfy the
kinetic bounds does not necessarily rule out certain parameter choices 
or quantization schemes, but only limits the allowed range 
where the approximation to ODEs can be employed. When the Hubble 
length becomes smaller than the fundamental discreteness scale,
$a_i$, the effective ODEs describing the cosmic dynamics 
become invalid and must be replaced by the full quantum equations,
which are a difference equation for the wave function in the case of
loop quantum cosmology \cite{bojowald02a}. It is possible that in this 
framework both the {\sc Ham} and {\sc Fried} schemes would 
become equally viable at a phenomenological level.

Considering values $l < 3/4$ does not alter the qualitative behavior
of the dynamics. In general, for a given initial kinetic energy, this
leads to an increase in the lowest value of $j$ consistent with
successful inflation in {\sc Fried} quantization, since the
semi--classical phase must last longer. Conversely, when $l >3/4$,
successful inflation is possible for lower initial kinetic energies
and lower values of $j$.  For {\sc Ham} quantization, reducing $l$
makes it harder to satisfy the horizon and kinetic bounds
simultaneously. The kinetic energy of the field increases less rapidly
for $l \rightarrow 0$ and it might therefore be expected that it would
be easier to satisfy the kinetic bound. However, in this case the
field lacks the kinetic energy needed to reach a sufficiently high
value after the semi-classical era. Thus, the phenomenological
indications for $l$ are in agreement with the expectations from the
full theory which lead to $l\geq 1/2$.

In fact, the results allow us to draw further lessons for the full
theory. As discussed in \S 2.2, {\sc Ham} is analogous to the
full quantization procedure whose dynamics is governed by a
constraint, while {\sc Fried} does not have a full analog. From the
phenomenological point of view we can separate the range of parameters
for different quantizations into three distinct domains: (i) values
which do not lead to sufficient slow-roll inflation; (iia) values
which lead to sufficient inflation in such a way that the scalar field
reaches a turning point around $3l_{\rm Pl}^{-1}$; and (iib) values
leading to sufficient inflation with a turning point much higher than
$3l_{\rm Pl}^{-1}$. As we have seen, {\sc Fried} is phenomenologically
more robust to ambiguities and most cases fall into class (iib). On
the other hand, {\sc Ham} is less robust but has most realisations in
class (iia) which are more likely to lead to observable effects by putting the
maximal inflaton value just at the borderline for sufficient inflation. This
may indicate that observations of full loop quantum gravity are indeed
within reach.

As for $j$, the lower bounds are larger than unity, as expected, but not
unreasonably large. Thus, the scenario appears realistic and does not
require fine tuning, which is also a consequence of the weak
logarithmic dependence of $\phi_{\rm max}$ on $j$.

A key open question in LQC concerns 
the evolution of scalar (density) and tensor (gravitational wave) 
perturbations generated during the semi--classical epoch. 
Although standard perturbation theory is unstable 
in this regime \cite{tsm03}, it is possible that  
these perturbations 
may imprint characteristic signatures 
of this phase on the CMB power spectrum. Indeed, 
there have recently been a number of suggestions for explaining the 
apparent suppression of the CMB power spectrum on large angular 
scales through quantum gravity effects, including LQC \cite{tsm03}, 
non--commutative geometry \cite{roy1} and higher--order curvature terms 
in string--inspired models \cite{shinji}. It is important, therefore, to 
investigate whether observations will be able to discriminate 
between these different possibilities. 
Finally, it would also be 
interesting to extend the analysis of this work to the regime 
of the quantum difference equations of the full theory and 
to
investigate whether any observable effects for the 
different quantization schemes can be uncovered.  

\vspace{.3in}
\centerline{\bf Acknowledgments}
\vspace{.3in}
We are grateful to Roy Maartens for discussions. DM is supported 
by a PPARC studentship. PS thanks Queen Mary,
University of London for warm hospitality during the initial stages of
this work. He is supported by a CSIR grant.

\begin{appendix}
\section{Inverse metric in the full theory}

Loop quantum gravity is based on a classical formulation of general
relativity using Ashtekar variables. These variable are the densitized triad
$E^a_i$, related to the spatial metric via
\[
 E^a_iE^b_i=q^{ab}\det q
\]
and the connection $A_a^i=\Gamma_a^i-\gamma K_a^i$ forming a conjugate
pair. Here, $\Gamma_a^i$ is the spin connection of the triad,
$K_a^i$ the extrinsic curvature, and $\gamma$ the Barbero--Immirzi
parameter.

The quantization then proceeds by using as basic quantities holonomies
of the connection, $h_e(A)={\cal P}\exp\int_e A$ for curves $e$ in
space, and fluxes computed from the triad, $F_S(E)=\int_SE$ for
spatial surfaces $S$. The important feature of these expressions is
that they are defined in a background independent way, i.e., we do not
need to choose a metric to define the integration. Yet, by integrating
the basic fields $E^a_i$, $A_a^i$ are smeared which renders the
quantization well-defined.

The background independence in particular requires the density weight
employed in the triad $E^a_i$. This means that the un-densitized triad
$e^a_i=E^a_i/\sqrt{|\det E|}$ and also the co-triad
$e_a^i=(e^{-1})_a^i$ are obtained by dividing by $\sqrt{|\det E|}$ and
inverting. In general this is ill-defined given that there may be 
possibly degenerate triads. Nevertheless, it is possible to compute the
co-triad, say, from the basic quantities in a well-defined way since
\cite{qsdi}
\begin{equation}\label{cotriad}
 e_a^i(x)=\frac{1}{4\pi G\gamma} \{A_a^i(x),\int d^3y \sqrt{|\det
E(y)|}\}
\end{equation}
where we have employed the Poisson bracket and integrated over an
arbitrary region containing $x$. This expression does not involve
inverses and can be quantized in a well-defined way using holonomies,
the volume operator, and turning the Poisson bracket into a
commutator. Such classical identities lie at the heart of quantizing
expressions as matter Hamiltonians in the full theory
\cite{thiemann_matter} and the geometrical density in isotropic
models.

For a scalar field Hamiltonian we need a quantization of $|\det E|^{-1/2}$,
which reduces to $a^{-3}$ in an isotropic model. The identity
(\ref{cotriad}) cannot be used immediately, but we can make another
reformulation:
\begin{eqnarray*}
 \frac{1}{\sqrt{|\det E|}}=\frac{\det e}{\det E}=
\frac{\frac{1}{6}\epsilon^{abc}\epsilon_{ijk} e_a^ie_b^je_c^k}{\det
E}\\
= \frac{1}{6}\epsilon^{abc}\epsilon_{ijk} (4\pi G\gamma)^3
\{A_a^i,V^{1/3}\} \{A_b^j,V^{1/3}\} \{A_c^j,V^{1/3}\}
\end{eqnarray*}

\vspace{.1in}
\noindent by writing $V$ for $\int d^3y \sqrt{|\det E(y)|}$. 
This expression can
now be embedded in the full matter Hamiltonian and quantized in a
background independent way (see \cite{thiemann_matter} for
details). Upon inserting isotropic variables and expressing the
connection via holonomies, one can easily check that one obtains
(\ref{dens}) with $l=1/2$ (in the Poisson brackets we have simply
$a=V^{1/3}$ in both cases).

Even taking into account the need for a background independent
quantization, there is still some freedom in the full theory. For 
example, we can
multiply by any positive integer power of $\det e$, such that
\begin{eqnarray*}
  \frac{1}{\sqrt{|\det E|}}=\frac{(\det e)^k}{(\det E)^{(k+1)/2}}=
\left(\frac{1}{6}\epsilon^{abc}\epsilon_{ijk} (4\pi
G\gamma)^3\right.\\
\times 
\{A_a^i,V^{(2k-1)/3k}\} \{A_b^j,V^{(2k-1)/3k}\}
\{A_c^j,V^{(2k-1)/3k}\}\Bigr)^k
\end{eqnarray*}
with a discrete rather than a continuous set of the ambiguity
parameter. When evaluating in isotropic variables, we obtain the form
(\ref{dens}) with $l_k=1-(2k)^{-1}\geq1/2$. The value $l_2=3/4$
results from using $\det E$ to multiply and this seems the most natural
choice when we keep in mind that $E$, rather than $e$, 
is the basic geometrical
variable underlying loop quantum gravity. The analogous reformulation
of $1/\sqrt{|\det E|}$ has been used in \cite{thiemann_matter} for a
full scalar Hamiltonian (see \S 3.3 of that work). Moreover, 
since $l_2$ is the
midpoint of the set spanned by the values $l_k$, it should
give representative results.

\end{appendix}

\end{document}